\documentclass{article}
\usepackage{ijcai19}

\usepackage{times}
\usepackage{soul}
\usepackage{url}
\usepackage[draft]{hyperref}
\usepackage[utf8]{inputenc}
\usepackage[small]{caption}
\usepackage{graphicx}
\usepackage{amsmath}
\usepackage{booktabs}
\usepackage{amssymb}
\usepackage{subfigure}
\usepackage{ctable, multirow}
\usepackage{colortbl}
\usepackage[boxed]{algorithm2e}

\usepackage{hhline}
\usepackage{amsthm}
\usepackage{pifont}
\newcommand{\cmark}{\ding{51}}%
\newcommand{\xmark}{\ding{53}}%

\theoremstyle{definition}
\newtheorem{definition}{Definition}[section]

\graphicspath{{./}{Figures/}}
\makeatletter

\newcommand{\Rmnum}[1]{\expandafter\@slowromancap\romannumeral #1@}
\newcommand{\G}{\mathcal{G}}
\newcolumntype{Y}{>{\centering\arraybackslash}X} 
\definecolor{Gray}{gray}{0.9}

\title{Improved Deep Embeddings for Inferencing with Multi-Layered Networks}

\author{
	Huan Song$^1$\And
	Jayaraman J. Thiagarajan$^2$\footnote{This work was performed under the auspices of the U.S. Department of Energy by Lawrence Livermore National Laboratory under Contract DE- AC52-07NA27344.}\\
	\affiliations
	$^1$Bosch Research North America\\
	$^2$Lawrence Livermore National Labs\\
	\emails
	huan.song@us.bosch.com,
	jjayaram@llnl.gov
}

\begin{document}
\maketitle

\begin{abstract}
Inferencing with network data necessitates the mapping of its nodes into a vector space, where the relationships are preserved. However, with multi-layered networks, where multiple types of relationships exist for the same set of nodes, it is crucial to exploit the information shared between layers, in addition to the distinct aspects of each layer. In this paper, we propose a novel approach that first obtains node embeddings in all layers jointly via DeepWalk on a \textit{supra} graph, which allows interactions between layers, and then fine-tunes the embeddings to encourage cohesive structure in the latent space. With empirical studies in node classification, link prediction and multi-layered community detection, we show that the proposed approach outperforms existing single- and multi-layered network embedding algorithms on several benchmarks. In addition to effectively scaling to a large number of layers (tested up to $37$), our approach consistently produces highly modular community structure, even when compared to methods that directly optimize for the modularity function. 

\end{abstract}

\section{Introduction}
Modeling and inferencing with network data are essential to several application domains including social network analysis \cite{eagle2006reality}, recommendation systems \cite{rao2015collaborative} and healthcare \cite{fornito2013graph}. This broadly encompasses problems such as node classification, link prediction, community detection, influential node detection etc. They are typically solved by mapping the nodes into a latent low-dimensional vector space, and invoking appropriate machine learning techniques on the resulting embeddings, e.g. unsupervised clustering for community detection~\cite{ding2016node}. In its simplest form, such embeddings can be inferred via descriptors of the network structure -- e.g. the graph adjacency, the graph Laplacian or the modularity matrix~\cite{chen2014community,yang2016modularity}. While the earliest solutions relied on matrix decomposition for optimizing embeddings, more recently, deep graph autoencoders have been found to be more flexible~\cite{yang2016modularity,thiagarajan2016robust}. However, the lack of scalability has been a persistent limitation of these approaches. This motivated the design of state-of-the-art approaches such as DeepWalk~\cite{perozzi2014deepwalk} and Node2Vec~\cite{grover2016node2vec} that are based on a distributional hypothesis, popularly adopted in language modeling~\cite{harris1954distributional}, where co-occurrence of two nodes in short random walks implies a strong notion of semantic similarity. Supported by efficient optimization strategies such as negative sampling~\cite{mikolov2013distributed}, these methods can easily deal with very large-scale networks.

Recently, with the emergence of multi-layered networks, \textit{i.e.}, the same set of nodes with different relational structure in each layer~\cite{li2018multi}, obtaining concise embeddings that preserve the multi-view relationships has become crucial. In contrast to the single-layer case, node embedding techniques for multi-layered networks must accommodate the extraction of both layer-wise (e.g. community detection) as well as unified representations (e.g. node classification). Due to the inherent challenges in multi-layered networks including heterogeneity in the relationship types and varying levels of sparsity in different layers, straightforward extensions of single-layer methods are ineffective~\cite{zhang2018scalable}. Finally, scalability with respect to increasing number of layers, is an important design criterion. 

In this paper, we develop a scalable embedding approach that first performs a DeepWalk-style optimization directly on the multi-layered network (\texttt{M-DeepWalk}), and utilizes a refinement strategy to further fine-tune the embeddings by encouraging cohesive community formation. We study the application of our approach to link prediction, node classification and community detection problems. Surprisingly, through the inclusion of multi-layer dependencies, we find that \texttt{M-DeepWalk} can already outperform state-of-the-art multi-layered embedding methods~\cite{zhang2018scalable} in link prediction. However, with more challenging tasks such as node classification and multi-layered community detection, there is room for improvement. The premise of using short random walks to infer the underlying semantic structure relies on the assumption that the networks are highly sparse and the node co-occurrences follow a power law. However, by allowing inter-layer edges, that assumption can be violated in cases where the semantics persist over even longer walks. We address this limitation by including a \textit{refinement} step, where the embeddings from \texttt{M-DeepWalk} are fine-tuned to produce more cohesive communities. Even with network datasets containing as many as $37$ layers, we show that the proposed approach produces highly modular community structure, when compared to existing multi-layered community detection techniques.

\section{Problem Definition}
\label{sec:problem}
We represent a single-layer undirected, unweighted network as $\mathcal{G} = (\mathcal{V}, \mathcal{E})$, where $\mathcal{V}$ denotes the set of nodes with cardinality $|\mathcal{V}| = N$, and $\mathcal{E}$ denotes the set of edges. The goal of an embedding technique is to generate latent representations for the nodes, $\mathbf{X} \in \mathbb{R}^{N \times d}$, where $d$ is the desired embedding dimensionality. Correspondingly, we denote a multi-layered network using a set of $L$ inter-dependent networks $\mathcal{G}^{(l)} = (\mathcal{V}^{(l)}, \mathcal{E}^{(l)}), \text{for } l = 1,\dots,L$, where there exists a node mapping $\mathcal{M}$ between every pair of layers to indicate which nodes in one network correspond to nodes in the other. Note that we expect different layers to contain common nodes, but do not require $\mathcal{V}^{(l)}$ to be the same. The complete set of nodes in the multi-layered network is denoted as $\mathcal{V}=\cup_{l=1}^{L}\mathcal{V}^{(l)}$, where $\cup$ refers to union. Now, we define the inferencing tasks considered in this paper, which require the extraction of multi-layered network embeddings.

\theoremstyle{definition}
\begin{definition}{(\textit{Node classification})}
Given a multi-layered network $\{(\mathcal{V}^{(l)}, \mathcal{E}^{(l)})\}_{l=1}^L$ and the semantic labels $\mathcal{Y}_{lab}$ for a subset of nodes $\mathcal{V}_{lab} \subset \mathcal{V}$, where each $y \in \mathcal{Y}_{lab}$ assumes one of the $K$ pre-defined classes, predict labels for each of the nodes in the set $v \in \mathcal{V} \setminus \mathcal{V}_{lab}$.
\end{definition}

\theoremstyle{definition}
\begin{definition}{(\textit{Link prediction})}
	Given a multi-layered network $\{(\mathcal{V}^{(l)}, \mathcal{E}^{(l)})\}_{l=1}^L$, in a layer $i$, estimate the likelihood $p(e | l = i)$ for an unobserved edge $e \in \mathcal{E} \setminus \mathcal{E}^{(i)}$ to exist, where $\mathcal{E}$ is the superset of all possible edges in layer $i$.
\end{definition}

\theoremstyle{definition}
\begin{definition}{(\textit{Multi-layered community detection})}
	Given a multi-layered network $\{(\mathcal{V}^{(l)}, \mathcal{E}^{(l)})\}_{l=1}^L$, identify coherent communities denoted by the sets $\mathcal{C}_1, \cdots, \mathcal{C}_K$, such that each node $i$ in every layer $l$ is assigned to one of the communities, $\arg \max_k \mathcal{A}(v_i^{(l)}, \mathcal{C}_k)$. Here, $\mathcal{A}$ denotes a similarity function.
\end{definition}

\section{Proposed Approach}
\begin{figure}[t]
	\centering
	\includegraphics[width=1\linewidth]{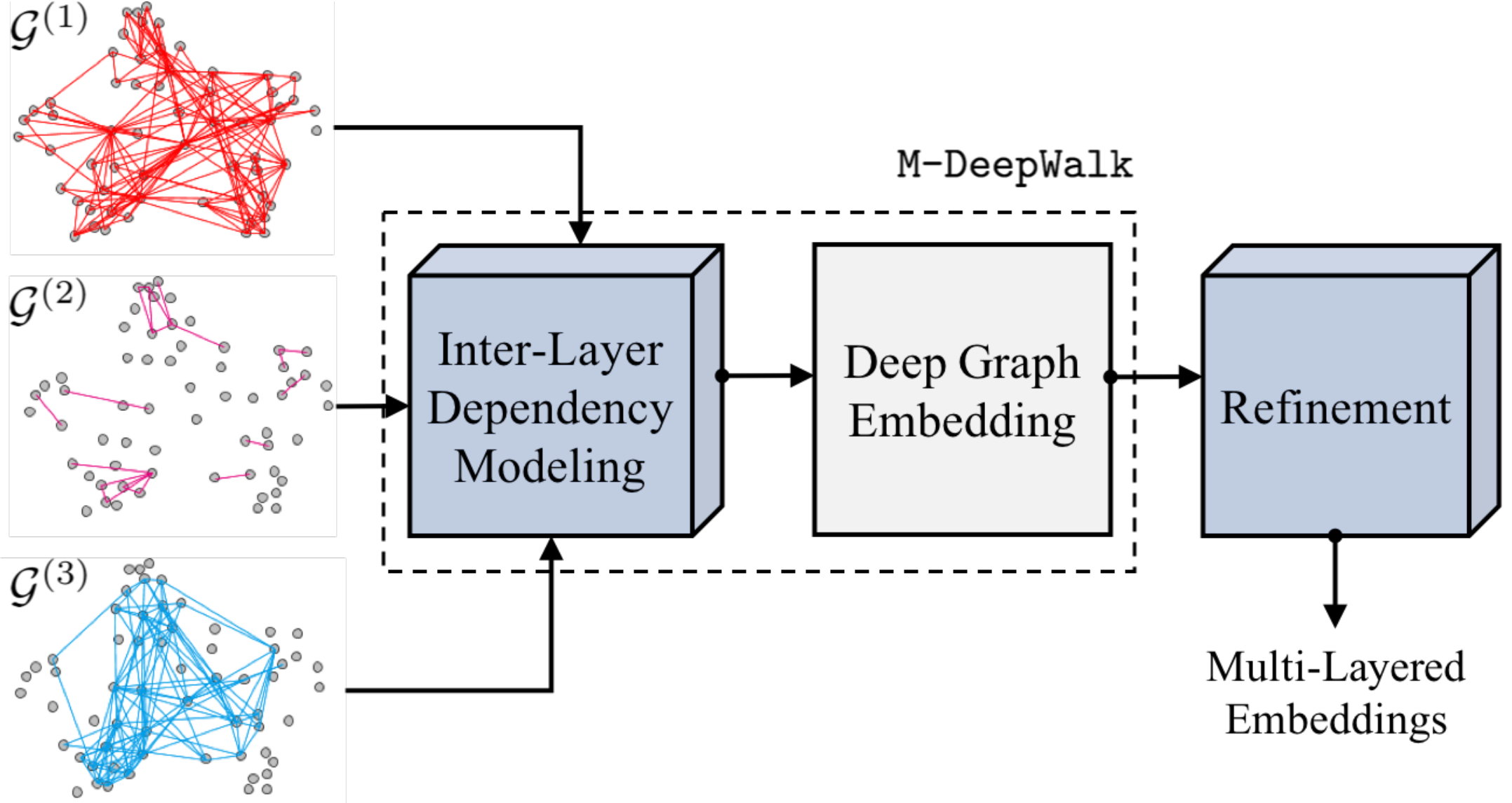}
	\caption{Proposed approach for extracting node embeddings from multi-layered networks.}
	\label{fig:arch}
	\vspace{-0.1in}
\end{figure}





As illustrated in Figure \ref{fig:arch}, our approach takes as input a multi-layered network, and infers $d$-dimensional latent representations for each of the nodes in $\mathcal{V}=\cup_{l=1}^{L}\mathcal{V}^{(l)}$. In contrast to state-of-the-art approaches~\cite{liu2017principled,li2018multi}, we do not simultaneously optimize for common representations that are applicable to all layers and local representations that preserve the layer-specific network structure. Instead, we construct a multi-layer supra graph $\mathcal{G}^{sup}$, that explicitly models inter-layer dependencies, and obtain embeddings for each node in every layer using deep network embedding methods. While this process produces semantically meaningful embeddings, we show that it is beneficial to fine-tune the embeddings through refinement strategies that encourage cohesive community structure in the latent space.

\subsection{Node Embeddings using \texttt{M-DeepWalk}}

\begin{figure*}[t]
	\centering
	\subfigure[Construction of supra graph for \texttt{M-DeepWalk}.]{\includegraphics[width=0.47\linewidth]{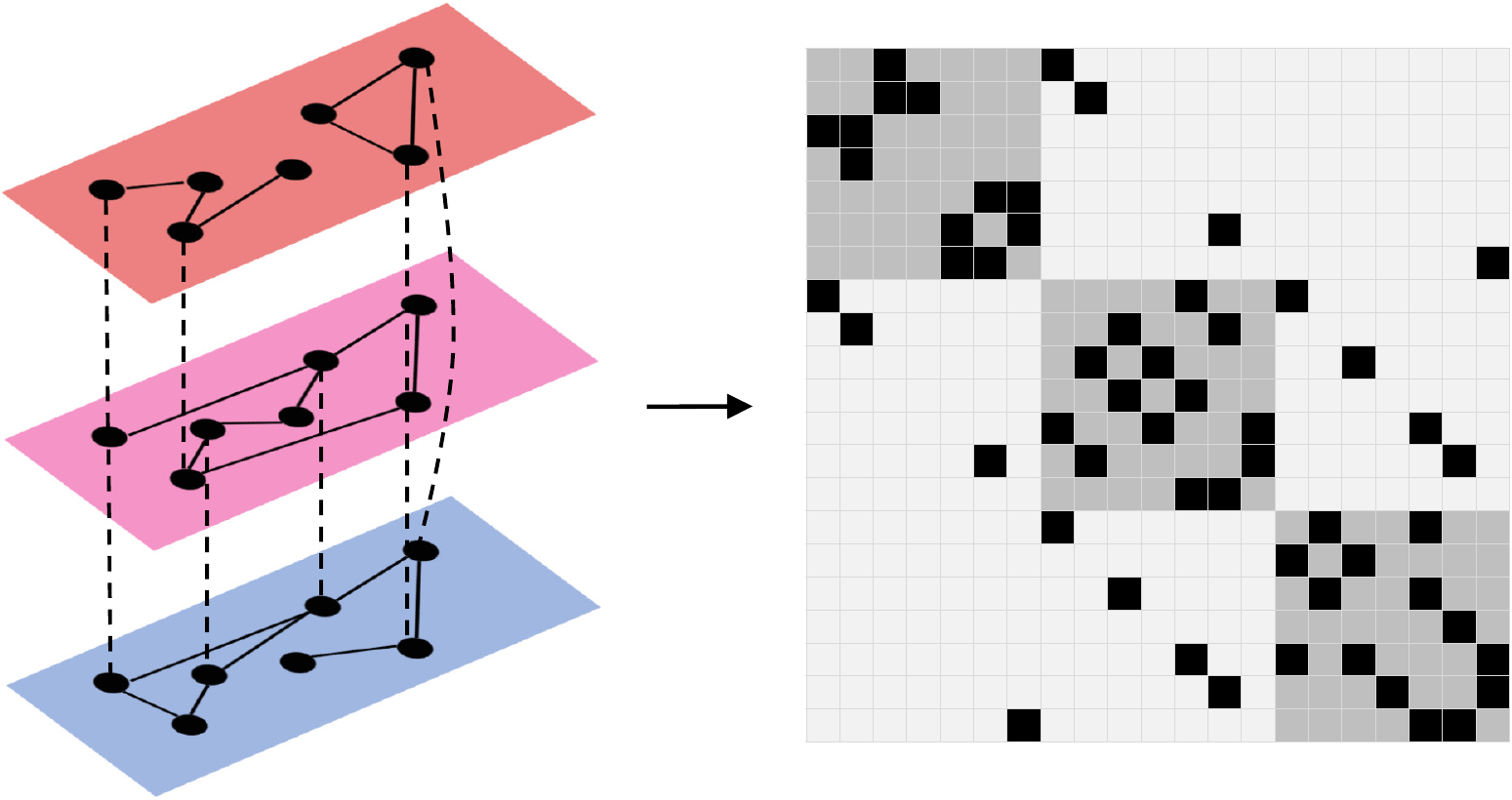}\label{fig:early}}
	\hfill
	\subfigure[Refining multi-layered embeddings from \texttt{M-DeepWalk} to produce cohesive structure.]{\includegraphics[width=0.47\linewidth]{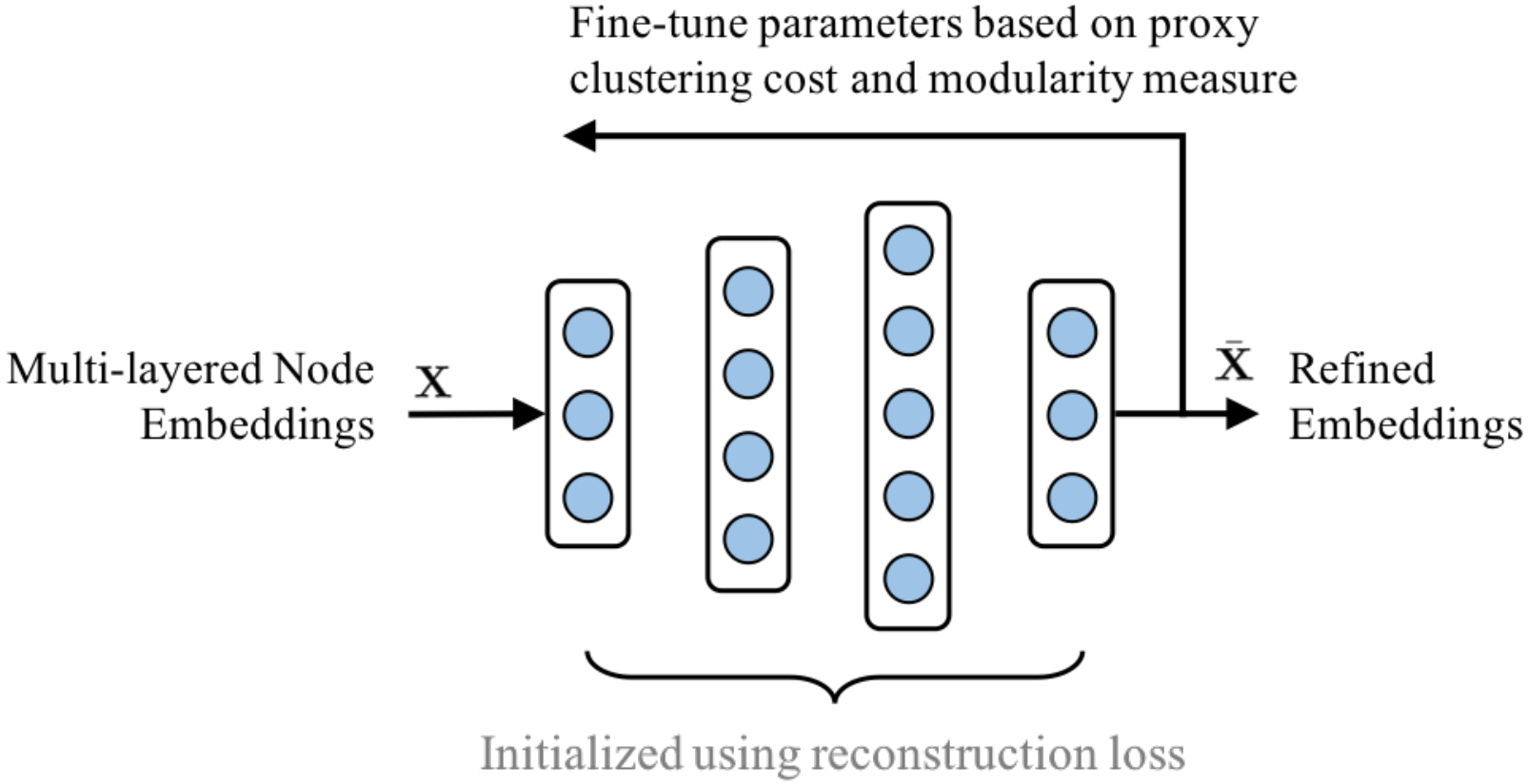}\label{fig:refine}}
	\caption{(a) We introduce edges between a node and its counterparts in different layers to exploit the shared local structures. Here, we illustrate the supra graph construction for an example case with $7$ nodes and $3$ layers. The solid lines denote the existing intra-layer edges and the dashed lines denote the virtual inter-layer edges. Similarly, while the block diagonals in the adjacency matrix correspond to individual layers, the off-diagonal entries encode the inter-layer edges; (b) We refine the latent embeddings to produce cohesive community structure, through the use of a \textit{proxy} clustering cost and modularity refinement.}
	\label{fig:early_and_refine}
\end{figure*} 

We aim to exploit the dependencies across different layers by creating inter-layer edges, wherein such dependencies can be encoded into the latent representations through deep network embedding techniques such as DeepWalk~\cite{perozzi2014deepwalk}. Broadly, structure discovery in multi-layered networks can be characterized using two scenarios: (a) a semantically coherent group can comprise nodes from all or a subset of the layers; and (b) a semantic concept can be discovered only within a specific layer and it is not apparent in other layers. This motivates the design of an embedding technique that respects the following constraints: For scenario (a), it is necessary to exploit the information flow across layers, whereas for the latter scenario, the latent embeddings should not be confused by the dependencies in other layers. In order to achieve this, we introduce inter-layer edges based on the similarities between local neighborhood structure. For a pair of nodes $v_i^{(l)}$ and $v_j^{(m)}$ from the layers $l$ and $m$ respectively, we introduce inter-layer edges as follows:
\begin{equation}
\label{eq:earlyweight}
\begin{gathered}
e_{ij}^{(l,m)}=0, \text{if } i\neq j, \\
e_{ij}^{(l,m)}=\frac{|\mathcal{N}_i^{(l)}\cap \mathcal{N}_j^{(m)}|}{|\mathcal{N}_i^{(l)}\cup \mathcal{N}_j^{(m)}|}, \text{if } i=j
\end{gathered} 
\end{equation}
where $e_{ij}^{(l,m)}$ denotes the edge weight between the nodes $v_i^{(l)}$ and $v_j^{(m)}$, and $\mathcal{N}_i^{(l)}, \mathcal{N}_j^{(m)}$ are the neighborhoods of $v_i^{(l)}$ and $v_j^{(m)}$ respectively. Note that, the edge weight is computed as the Jaccard coefficient of the agreement in neighborhood lists between the two layers. This ensures that each node in a layer is connected only to its counterpart in another layer, and similar local neighborhoods is a strong indicator of a shared semantic structure. Upon computation of the edge weights, we convert $e_{ij}^{(k,l)}$ to binary connectivities using a specified threshold. The multi-layered network with the virtual inter-layer edges is referred as a \textit{supra graph}, $\G^{sup}$ (Figure \ref{fig:early}).

DeepWalk draws analogy between node sequences generated by short random walks on networks and sentences in a document corpus. Let us consider a simple metric walk $\mathcal{W}_t$ at step $t$ in a single-layer network $\mathcal{G}$, which is rooted at the node $v_i$. The transition probability between nodes $v_i$ and $v_j$ is $P(\mathcal{W}_{t+1} = v_j | \mathcal{W}_t = v_i) = h(\|\mathbf{x}_i - \mathbf{x}_j\|_2 / \sigma)$, where $\|\mathbf{x}_i - \mathbf{x}_j\|_2$ measures the closeness between two nodes in the latent space and $h$ is a linking function that connects the node similarity to the actual co-occurrence probability. With appropriate choice of the walk length, the true metric can be recovered accurately from the co-occurrence statistics constructed using random walks. Denoting a $S$-step random walk on the network as $(v_0, v_1,\dots,v_{S-1})$, we attempt to obtain a vector space that can recover the likelihood of observing a node given its context, i.e., $P(v_S | v_0, v_1,\dots,v_{S-1})$ over several random walks. In the proposed approach, we utilize DeepWalk to learn embeddings for the $|\mathcal{V}|$-nodes in $\G^{sup}$, and this is referred as \texttt{M-DeepWalk}. 

%
%
%
%
%

\subsection{Refinement Step}

The success of \texttt{M-DeepWalk} relies on the assumption that networks are sparse and short random walks reveal all necessary semantics. However, due to the introduction of inter-layer edges, and varying levels of sparsity in different layers, this assumption can be violated in cases where the semantic relationships persist over longer walks. In order to address this challenge and to produce cohesive structure in the latent space, we propose to refine the learned embeddings. More specifically, we employ a \textit{proxy} clustering cost, along with modularity based reassignment, to refine embeddings.

\noindent \textbf{Architecture:} We utilize a $4-$layer fully connected neural network, $\mathcal{H}$, for implementing the refinement step. Since the representations from \texttt{M-DeepWalk} contain significant topological information as abstracted from the raw network data, we found that a reasonably shallow neural network is sufficient. As shown in Figure \ref{fig:refine}, the inputs to the network are embeddings $\mathbf{X} \in \mathbb{R}^d$ from \texttt{M-DeepWalk} and the goal is to produce fine-tuned embeddings $\bar{\mathbf{X}}$ (also in $\mathbb{R}^d$).


\begin{table*}[t]
	\renewcommand{\arraystretch}{1.2}
	\caption{Details for all datasets used in our experiments. Number of nodes and edges refers to the entire multi-layered network.}
	\begin{tabular}{c|c|c|c|c|c|c}
		\hline
		\rowcolor{Gray}
		{\textbf{Dataset}} & \textbf{\# Nodes} & \textbf{\# Edges} & \textbf{\# Layers} & \textbf{Node Classification} & {\textbf{Link Prediction}} & {\textbf{Community Detection}} \\ \hline \hline
		{Leskovec-Ng}      & 214               & 536               & 4                  &  \cmark &       \cmark  &   \xmark \\ 
		{Reinnovation}     & 1728              & 8,460             & 12                 &  \cmark &       \cmark  &   \xmark \\  
		{Congress Votes}   & 6672              & 710,717           & 16                  &  \cmark &       \cmark  &   \xmark \\ 
		{Mammography}      & 3844              & 1,562,814         & 4                  &  \cmark &       \cmark  &   \xmark \\ 
		Balance Scale                          & 2500              & 312,500           & 4      &  \cmark &       \cmark  &   \xmark \\ 
		Vickers                                & 87                & 740               & 3      &  \xmark &       \cmark  &   \xmark \\ 
		LAZEGA                                 & 211               & 2,571             & 3  &  \xmark &       \cmark  &   \cmark \\ 
		CELEGANS                               & 791               & 5,863             & 3 &  \xmark &       \cmark  &   \cmark \\
		EU Airlines                            & 2034              & 3,588             & 37 &  \xmark &       \xmark  &   \cmark \\  \hline
	\end{tabular}
\label{tab:datasets}
\vspace{-0.1in}
\end{table*}

\noindent \textbf{Proxy Clustering Loss:} We initialize the model $\mathcal{H}$ to an autoencoder, which minimizes a reconstruction objective (mean squared error).  Since it is not straightforward to measure how cohesive the resulting structure is, and actually use it to update $\mathcal{H}$, we resort to a \textit{proxy} clustering objective similar to~\cite{xie2016unsupervised}, \textit{i.e.}, produce low-entropy cluster assignment distributions. Assuming that there are $K$ clusters of nodes in the latent space, we first perform $k$-means clustering on $\bar{\mathbf{X}}$, and calculate the likelihood for a node to be assigned to each of the clusters as:
\begin{equation}
\label{eq:q}
q_{ik}=\frac{\bigg(1+\lVert \bar{\mathbf{x}}_i-\mu_k \rVert ^2/\alpha\bigg)^{-\frac{\alpha+1}{2}}}{\mathlarger{\sum}_{k'=1}^K \bigg(1+\lVert \bar{\mathbf{x}}_i-\mu_{k'} \rVert ^2/\alpha\bigg)^{-\frac{\alpha+1}{2}}},
\end{equation}where $\mu_k$ is the centroid for the $k^{th}$ cluster and $\alpha = 1$. Note, the layer index $l$ for each node is omitted for clarity, though the estimation is carried out for every $v_i^{(l)}$. Given the probabilities from (\ref{eq:q}), we iteratively refine the assignments by learning from regions of high confidence. First, we construct a target distribution for each node:
\begin{equation}
p_{ik}=\frac{q_{ik}^2/f_k}{\sum_{k'} q_{ik'}^2 / f_{k'}}
\label{eqn:aux}
\end{equation}which is a reduced-entropy variant of $q_{i}$. Here, $f_k = \sum_k q_{ik}$ are the soft cluster frequencies. We treat this target as a pseudo-supervision to adjust parameters of $\mathcal{H}$, such that each node is assigned to one of the clusters with high certainty:
\begin{equation}
\label{eq:L}
L=KL(P||Q)=\sum_i\sum_k p_{ik}\log(\frac{p_{ik}}{q_{ik}}).
\end{equation}Interestingly, this is similar to discriminative clustering methods \cite{ye2008discriminative} that iteratively perform clustering and supervised dimensionality reduction based on the current clustering, for maximizing separability. After this step, the algorithm updates $q_{ik}$ again according to eq. (\ref{eq:q}) and repeats this procedure until convergence, in terms of cluster assignment.

\noindent \textbf{Modularity-Driven Refinement:} Though the proxy clustering cost can be effective in producing cohesive structure, it relies heavily on the quality of the initial clustering. Hence, we propose to utilize the multi-slice modularity from the network to produce reliable embeddings. During each iteration of the refinement process, we move samples which have low modularity contribution to other communities, such that the modularity gain is maximized~\cite{duch2005community}. 

In general, the modularity function~\cite{newman2006finding} of a network is defined as the difference between the number of edges within cohesive communities and the expected number of edges over all pairs of nodes in a network. Formally,

\begin{equation}
\label{eq:modularity_single}
Q =\frac{1}{2r}\sum_{i,j}\bigg(e_{ij}-\frac{n_in_j}{2r}\bigg)\delta(c_i,c_j)
\end{equation}where $e_{ij}$ denotes the connection between nodes $v_i$ and $v_j$, $n_i,n_j$ denote the degrees for node $v_i$ and $v_j$ respectively, $r=\frac{1}{2}\sum_in_i$ is the total number of edges and $\delta(\cdot)$ is the Kronecker delta function which equals one only when the community memberships for nodes $v_i$ and $v_j$, namely $c_i$ and $c_j$, are the same. Extending this to the case of multi-layered networks provides the multi-slice modularity~\cite{mucha2010community}:
\small
\begin{align}
\label{eq:modularity_multi}
Q_{\text{multi}}=\frac{1}{2\mu}\sum_{i,j}\sum_{l,m}\biggl[&\biggl( e_{ij}^{(l)}-\gamma^{(l)}\frac{n_i^{(l)} n_j^{(l)}}{2r_l}\biggr)\delta(l,m) \nonumber \\
&+\delta(i,j)\sigma_j^{(l,m)}\biggr]\delta(c_i^{(l)},c_j^{(m)}),
\end{align}\normalsize where $\mu$ is a normalization factor, $\gamma^{(l)}$ is the resolution parameter for layer $l$, $\sigma_j^{(l,m)}$ is the coupling parameter between a pair of corresponding nodes in different layers, i.e., $v_j^{(l)}$ and $v_j^{(m)}$, and the other parameters are direct extensions from (\ref{eq:modularity_single}) to each layer $l$. Rewriting (\ref{eq:modularity_multi}) to describe the contribution of each node to the modularity function:
\begin{table*}[t]
	\centering
	\renewcommand{\arraystretch}{1.2}
	\caption{\textit{Node Classification} - Accuracy (\%) of the proposed approach in predicting node labels, when compared to baseline methods.}
	\begin{tabular}{c|c|c|c|c|c}
		\hline
		
		& \multicolumn{5}{c}{\textbf{Dataset}}                                                         \\ \cline{2-6} 
		
		\multirow{-2}{*}{\textbf{Method}} & \textbf{Leskovec-Ng} & \textbf{Reinnovation} & \textbf{Congress Votes} & \textbf{Mammography} & \textbf{Balance Scale} \\ \hline \hline
		DeepWalk                         & 99.2                 & 74.7                  & 99.8                    & 80.6                 & 90.9                   \\ 
		LINE                            & 91.1                 & 69.1                  & 98.0                      & 80.9                 & 78.5                   \\ 
		Node2Vec                         & 96.3                 & 76.0                   & 92.4                    & 80.2                 & 89.3                   \\ 
		PMNE                             & 94.5                 & 77.4                  & 98.4                    & 78.5                 & 91.1                   \\ 
		MNE                             & 92.4                 & 75.0                    & -                       &  74.3                    & 82.4                       \\ \hline \hline
		
		Proposed (w/o refine)        & 99.7                 & 76.0                    & \textbf{100}            & 81.3                 & 90.5                   \\ 
		
		Proposed (w/ refine)         & \textbf{100}         & \textbf{85.1}         & \textbf{100}            & \textbf{81.5}        & \textbf{92.1}          \\ \hline
	\end{tabular}
	\label{tab:node}
\end{table*}

\small
\begin{align}
\label{eq:fitness}
Q_{\text{multi}} & = \frac{1}{2\mu}\bigg[ \sum_l\sum_i\sum_{j\in \mathcal{C}_i^{(l)}} \bigg(e_{ij}^{(l)}-\gamma^{(l)}\frac{n_i^{(l)}n_j^{(l)}}{2r_l}\bigg) \nonumber \\
                 & \qquad\quad+\sum_l\sum_i\sum_{m \neq l}\sum_{j\in \mathcal{C}_i^{(l)}} \sigma_i^{(l,m)}\bigg] \nonumber \\
                 & = \frac{1}{2\mu}\bigg[ \sum_l\sum_i \bigg(n_{\mathcal{C}_i}-\gamma n_i^{(l)}\frac{r_{\mathcal{C}_i^{(l)}}}{r_l}\bigg)+\sum_l\sum_i\sigma n_{\mathcal{C}^{(l)}}\bigg] \nonumber \\
  				 & = \frac{1}{2\mu}\sum_l \sum_i\bigg(n_{\mathcal{C}_i}-\gamma n_i^{(l)}\frac{r_{\mathcal{C}_i^{(l)}}}{r_l}+\sigma n_{\mathcal{C}^{(l)}}\bigg)
\end{align}\normalsize This essentially extends the ``fitness score" in~\cite{duch2005community} to multi-layered networks. Here, we assumed that $\gamma_l,\sigma_i^{l,m}$ are constants for all nodes. We use $n_{\mathcal{C}_i}$ to denote the number of edges in layer $l$ that connect node $v_i^{(l)}$ to other nodes with the label $\mathcal{c}_i^{(l)}$. On the other hand, $n_{\mathcal{C}^{(l)}}$ denotes the number of counterpart nodes of $v_i^{(l)}$ which also share the label $\mathcal{c}_i^{(l)}$. The final summation term in eq. (\ref{eq:fitness}) defines the contribution from $v_i^{(l)}$ to the multi-slice modularity. Now, we measure the fitness score by normalizing the intra- and inter-layer contribution terms separately:
\begin{equation}
\label{eq:mod_contrib}
\lambda_i^{(l)}=\frac{n_{\mathcal{C}_i}}{n_i^{(l)}} -\gamma\frac{r_{\mathcal{C}_i^{(l)}}}{r_l}+\sigma\frac{n_{\mathcal{C}^{(l)}}}{n_i^{(l,m)}} 
\end{equation}where $n_i^{(l,m)}$ refers to the number of counterparts of $v_i^{(l)}$. Following the probabilistic selection process in \cite{duch2005community,boettcher2001s}, nodes in $\mathcal{V}$ are first sorted in ascending order based on their fitness scores, and then sampled according to the probability: $p(s)\propto s^{-\tau}$ where $s$ is the rank of node after sorting and $\tau\sim 1+\frac{1}{ln(|\mathcal{V}|)}$. Next, we move the sampled node $i'$ to community $k'$ which gives the largest gain of $Q_{\text{multi}}$. To reflect the updated assignment for the overall refinement process, we increase the community likelihoods $q_{i'k'}$ (eq. (\ref{eq:q})) by a constant $c$ and re-normalize the probabilities. Note that the change of community structure requires $\lambda_i^{(l)}$ to be re-calculated before next node is sampled. This sample and update process is repeated for a desired number of iterations. Subsequently, we carry out the calculation of eq. (\ref{eqn:aux}) and optimization in eq. (\ref{eq:L}). 


\section{Experiments}
\subsection{Datasets}
Table \ref{tab:datasets} lists the statistics of the datasets used in our experiments and indicates which tasks they were used to evaluate. 

\noindent \textbf{Leskovec-Ng}~\cite{chen2017multilayer}: This data describes $20$ year co-authorship information in $5$-year intervals, \textit{i.e.}, $4$-layers. In a layer, two researchers are connected if they co-authored in that $5$-year interval and the researcher's group affiliation is the label.

\noindent \textbf{Reinnovation}: This data describes the global innovation index for $144$ countries (nodes). This network contains $12$ layers encoding similarities in different sectors, e.g. infrastructure. The label at each node denotes the development level.

\noindent \textbf{Congress Votes}~\cite{schlimmer1987concept}: Based on the 1984 US congressional voting records, it includes votes from $435$ congressmen for $16$ bills, which results in a $16$-layered network. Each node is labeled as either a democrat or republican.

\noindent \textbf{UCI Mammography}~\cite{elter2007prediction}: This contains information about mammographic mass lesions from $961$ subjects, where the $4$ layers are constructed using attributes such as the BI-RADS assessment, density of the lesion etc.

\noindent \textbf{UCI Balance Scale}~\cite{Dua:2017}: This summarizes results from a psychological experiment, where $4$ attributes (left weight, the left distance, the right weight, and the right distance) were used to build the layers.

\noindent \textbf{Vickers}~\cite{vickers1981representing}: This models the social structure of $29$ students from Victoria, Australia. Each node is a student with gender as its label, and the three network layers are constructed based on student responses to a questionnaire.

\noindent \textbf{LAZEGA}~\cite{lazega2001collegial}: This multiplex network consists of $3$ relationship types (co-work, friendship and advice) between partners of a corporate partnership.

\noindent \textbf{C.ELEGANS}~\cite{chen2006wiring}: This multiplex network consists of $3$ layers describing synaptic junctions among a group of neurons.

\noindent \textbf{EU Airlines}~\cite{cardillo2013emergence}: In this air transportation multiplex network, each layer shows which cities have a direct flight between them with a certain European airline. There are $450$ cities (nodes) and $37$ airlines (layers) in total.


\begin{table*}[t]
	\centering
	\renewcommand{\arraystretch}{1.2}
	\caption{\textit{Link Prediction} - AUROC score of our approach in accepting positive links and rejecting negative links in multi-layered networks.}
	\begin{tabular}{c|c|c|c|c||c}
		
		\hline
		
		\multirow{2}{*}{\textbf{Dataset}} & \multicolumn{5}{c}{\textbf{Method}}                                                                                                                                                                        \\ \cline{2-6} 
		& \multicolumn{1}{l|}{\textbf{DeepWalk}} & \textbf{LINE}                  & \textbf{Node2Vec}              & \textbf{PMNE}                  &  \textbf{Proposed}  \\ \hline \hline
		Leskovec-Ng                       & \textbf{0.84}                          & 0.62 & 0.71 & 0.49 & \textbf{0.84}         \\ 
		Reinnovation                      & \textbf{0.99}                          & 0.78 & \textbf{0.99} & 0.78 & \textbf{0.99} \\ 
		Congress Votes                    & \textbf{1.0}                    & 0.99 & \textbf{1.0} & 0.79 & \textbf{1.0} \\ 
		Mammography                       & \textbf{1.0}                          & \textbf{1.0} & \textbf{1.0} & 0.77 & \textbf{1.0}  \\ 
		Balance Scale                     & \textbf{1.0}                    & \textbf{1.0} & \textbf{1.0} & 0.82 &  \textbf{1.0}  \\ 
		Vikcers                           & 0.83                          &   0.68                    &     0.76                  &     0.84                  &  \textbf{0.89}  \\ 
		LAZEGA                            & 0.88                          &   0.69                    &       0.8                &    0.82                  & \textbf{0.91}  \\ 
		C.ELEGANS                          & 0.93                          &    0.77                   &     0.89                  &      0.75                 &    \textbf{0.944}  \\  \hline
	\end{tabular}
	\label{tab:link}
\end{table*}

\begin{figure*}[t]
	\centering
	\subfigure[LAZEGA]{\includegraphics[width=.3\linewidth]{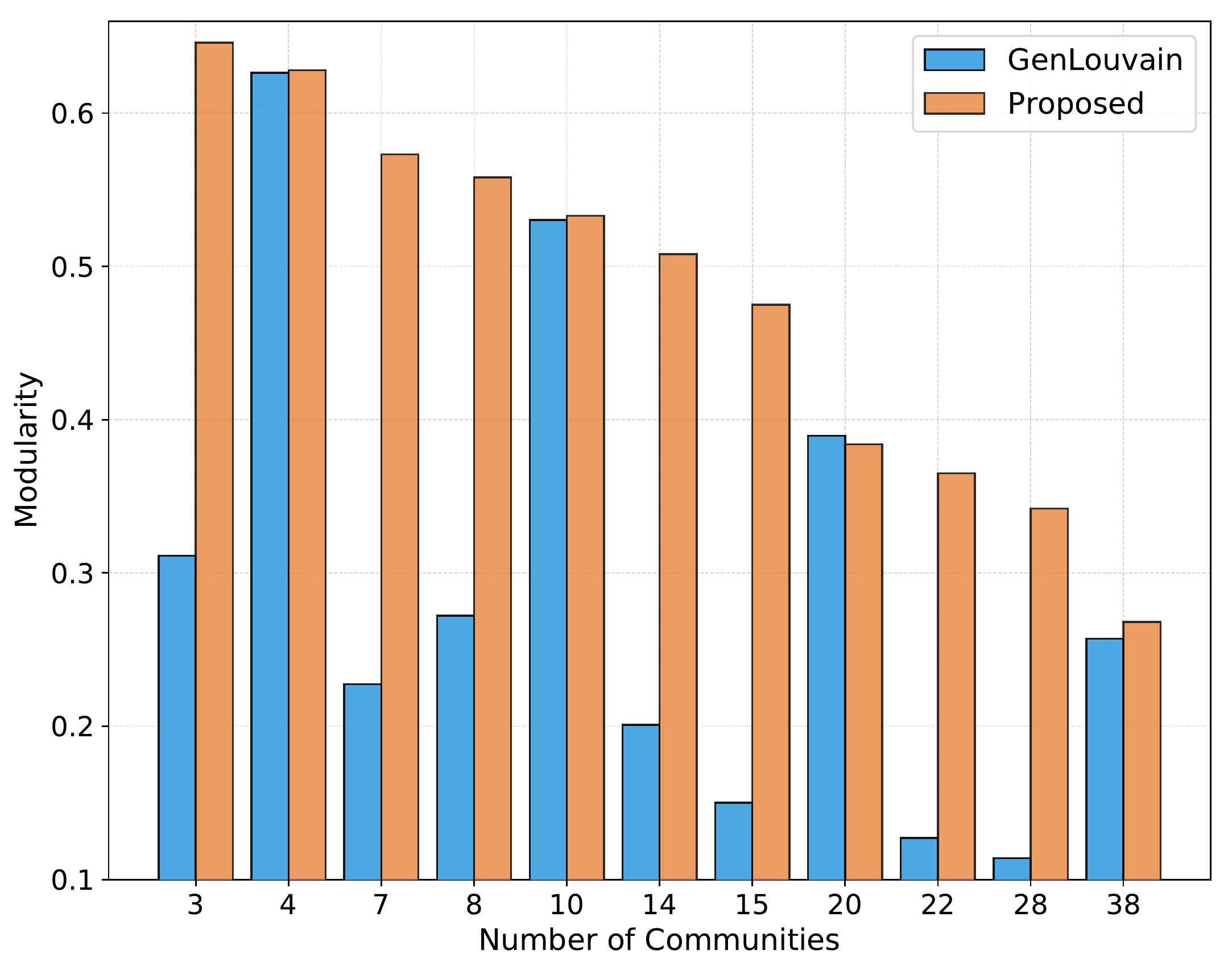}}
	\subfigure[C.ELEGANS]{\includegraphics[width=.3\linewidth]{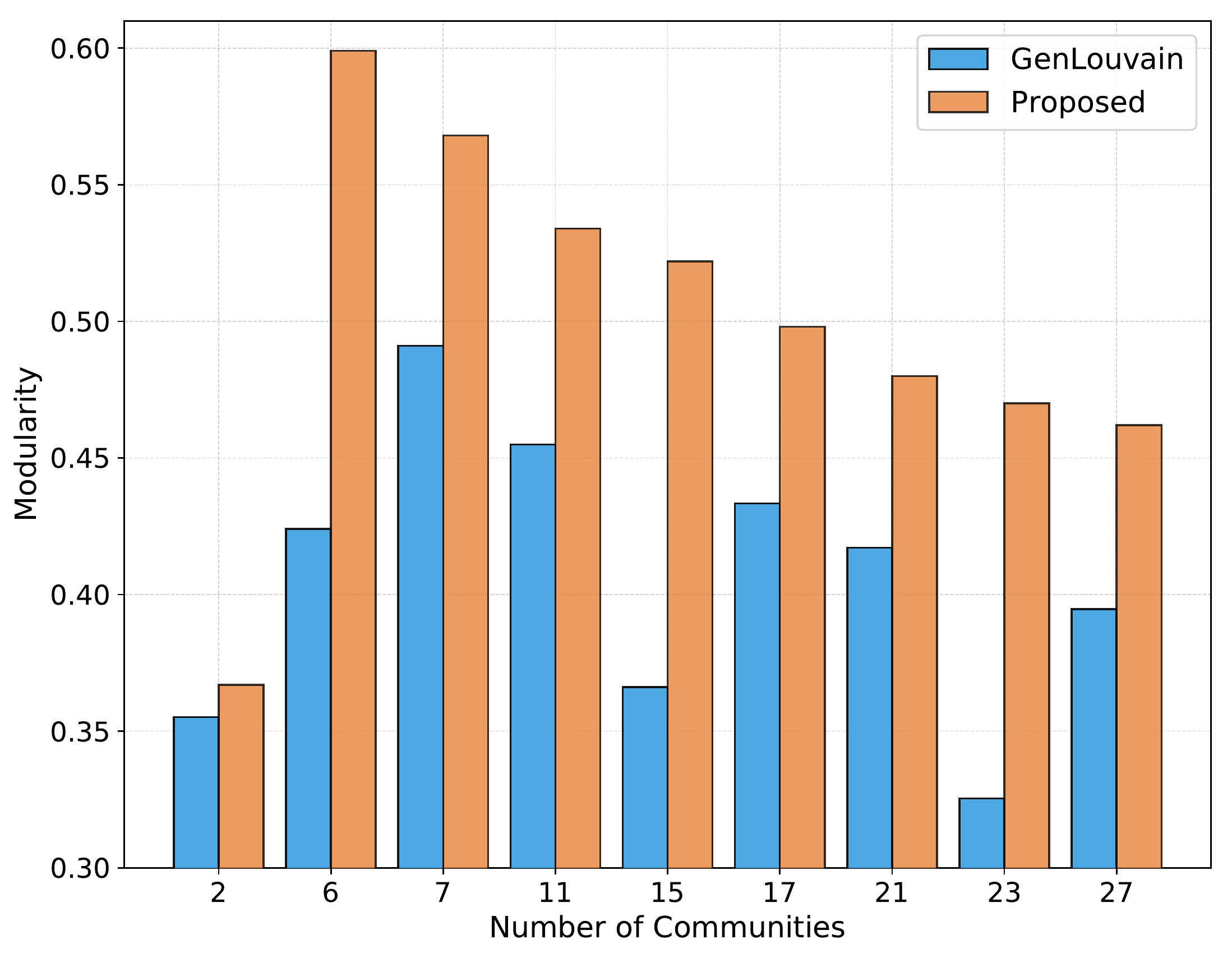}}
	\subfigure[EU Airlines]{\includegraphics[width=.3\linewidth]{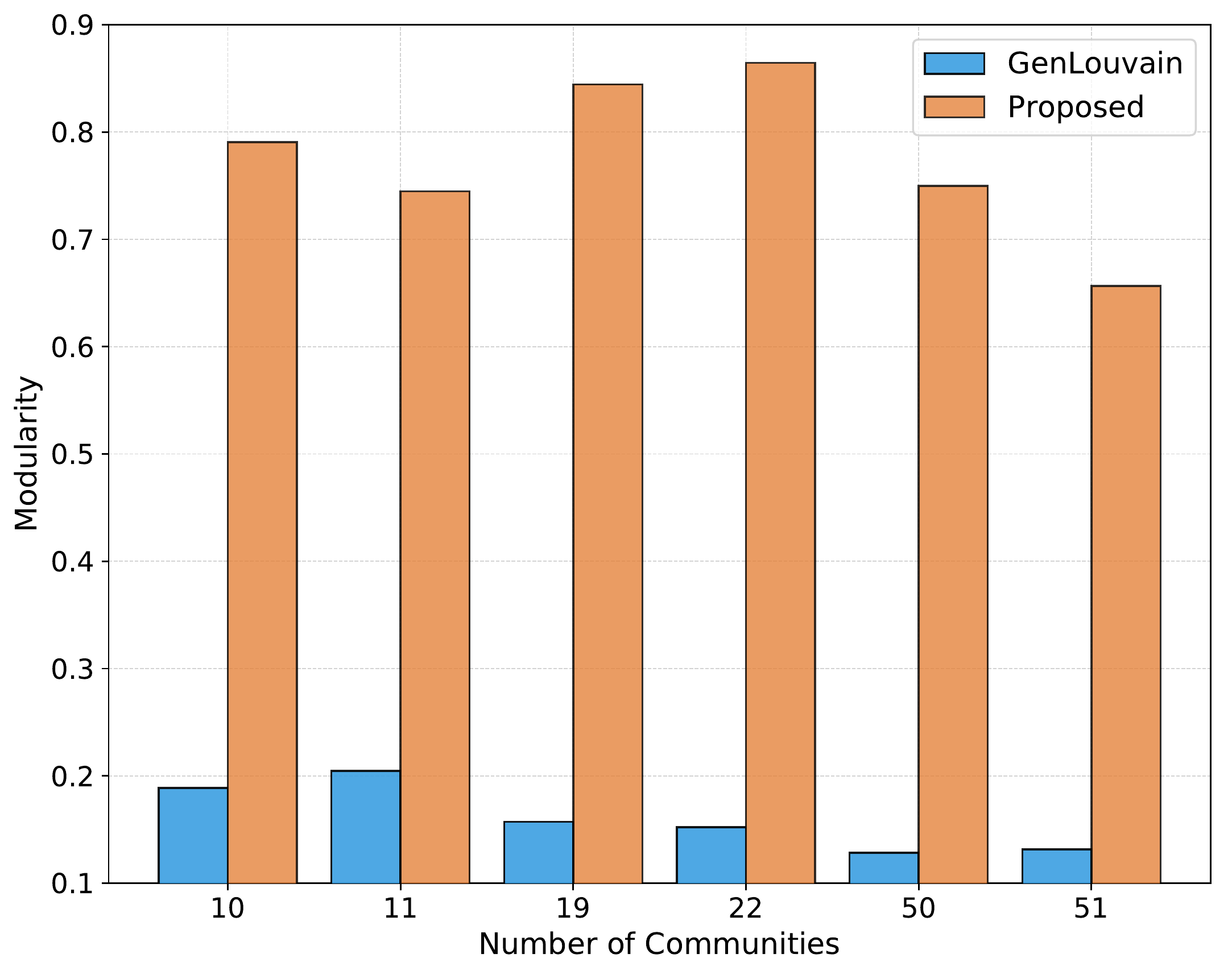}}
	\caption{\textit{Multi-layered community detection} - Modularity scores of our approach as compared to GenLouvain.}
	\label{fig:comm}
\end{figure*}

\subsection{Baseline Methods}
\noindent \textbf{DeepWalk}~\cite{perozzi2014deepwalk}: This first performs random walk on the network, and then uses the Skip-gram algorithm to infer embeddings.

\noindent \textbf{LINE}~\cite{tang2015line}: This improves over DeepWalk by adding a direct link fitting term to its cost function, and exploits higher-order relations via 2-hop neighborhoods.

\noindent \textbf{Node2Vec}~\cite{grover2016node2vec}: By parameterizing depth-first and breadth-first random walks in DeepWalk, this produces better embeddings for certain types of nodes.

\noindent \textbf{PMNE}~\cite{liu2017principled}: This employs different merge strategies to combine embeddings from each of the layers in a multi-layered network. In particular, we consider the results aggregation strategy, since it often outperforms other variants.

\noindent \textbf{MNE}~\cite{zhang2018scalable}: This recent embedding approach jointly optimizes for common embeddings that control information transfer while producing an additional representation to capture the distinct aspects of a layer.

\noindent \textbf{GenLouvain}~\cite{jutla2011generalized}: This modularity based community detection algorithm is very popular for extracting cohesive communities in multi-layered networks and produces state-of-the-art results on several datasets.

\subsection{Node Classification}
In this experiment, we assume that the true labels are observed for only a subset of the nodes and we predict the labels for the remaining nodes using a classifier trained on the node embeddings. More specifically, for each of the datasets we used a linear SVM classifier and the results reported are obtained using $3-$fold cross validation (with the smaller partition for training). The node classification accuracy was used as the evaluation metric. Table \ref{tab:node} shows the performance obtained using our proposed approach as well as the baseline methods. Note that, for the proposed approach, we show the results obtained with and without the refinement step. The first striking observation is that \texttt{M-DeepWalk} already produces highly competitive performance on all datasets, when compared to existing single-layer and multi-layer methods. Second, incorporating the proposed refinement strategy improves the performance further and thus produces state-of-the-art results. For example, with the Reinnovation dataset, our method achieves $10\%$ higher accuracy than the recent multiplex network embedding approach~\cite{zhang2018scalable}. Note, on the \textit{congress votes} data, the publicly released CPU-version of MNE either did not converge stably and could not produce valid embeddings.

\subsection{Link Prediction}
For each of the datasets, we randomly choose a subset of edges in each layer and remove them from the multi-layered network. Next, we extract node embeddings using the incomplete network, and estimate the likelihood for each of the missing edges. While the edges that were randomly removed are marked as positive, those edges which were not originally present are negative. We evaluate the link prediction performance by the algorithm's ability to accept positive links while rejecting the negative links. The likelihood is estimated as the cosine similarity between the two nodes corresponding to an edge. In our experiment, we performed a $5$-fold cross validation on the set of edges in each layer and report AUROC (area under the ROC curve) as the evaluation metric. In contrast to the node classification task, this task relies heavily on the local structure in each of the layers, and hence single-layer embeddings form a strong baseline. Surprisingly, with hyperparameter tuning (e.g. sufficient number of walks), simple baselines like DeepWalk are more competitive than observed in~\cite{zhang2018scalable}. As observed in Table~\ref{tab:link}, for $5$ out of the $9$ datasets, DeepWalk performs as well as more sophisticated methods without any additional inter-layer information. However, in cases where the information transfer from other layers is required to reliably predict a link, the proposed approach provides improvements, for example in Vickers, LAZEGA and C.ELELGANS datasets. Another observation is that including the refinement step did not provide apparent improvements in this task.

\subsection{Multi-layered Community Detection}
In the final experiment, we evaluate the proposed approach on multi-layered community detection, in terms of producing cohesive community structure. As defined in Section \ref{sec:problem}, this formulation can form communities that are comprised of nodes from multiple layers. The multiplex networks that we consider for this task are unlabeled, and hence we resort to unsupervised learning. For evaluation, we utilize the modularity metric. Most existing multi-layered network embedding approaches either do not explicitly optimize for modularity or constraint the community assignment across layers to be the same. Hence, we compare our results to the popular GenLouvain algorithm. The number of communities is an important parameter and for a comprehensive comparison, we generate different number of communities produced with GenLouvain, by varying the $\gamma$ and $\sigma$. For each of those cases, we performed $k$-means on the embeddings from our approach, with the same number of communities. The performance of the GenLouvain and the proposed approach are visualized in Figure \ref{fig:comm}. We observe that the performance of GenLouvain is highly sensitive to the number of communities chosen. In many cases, the algorithm fails to discover meaningful communities. On the contrary, our approach is consistently better and stable.
\section{Conclusions}
This paper presents a novel approach to extract node embeddings from multi-layered networks. While the locally adaptive inter-layer edge construction facilitates complex dependency modeling for \texttt{M-DeepWalk}, the refinement stage produces more cohesive structure in the latent space. Empirical studies demonstrate that the proposed approach is highly effective in a variety of tasks including link prediction, node classification and multi-layered community detection. In addition to outperforming existing single- and multi-layer embedding methods, we showed that the proposed approach scales to a large number of layers and produces highly effective node embeddings. 

\small
\bibliographystyle{named}
\bibliography{paper}

\end{document}